\documentclass[proceedings]{JHEP3}

\PrHEP{PrHEP hep2001}			
\conference{International Europhysics Conference on HEP}		

\usepackage{epsfig}                   

\newcommand{\alphapom}{\alpha_{_{\rm I\!P}}}

\newcommand{\av}[1]{\mbox{$ \langle #1 \rangle $}}
\newcommand{\xpom}{x_{_{I\!\!P}}}
\newcommand{\mx}{M_{_X}}

\newcommand{\lapprox}{\stackrel{<}{_{\sim}}}

\title{Measurements of the Diffractive Structure Function 
{\boldmath $F_2^{D(3)} (\beta, Q^2, \xpom)$} at HERA.}

\author{\speaker{Paul Newman}\thanks{Supported by the UK Particle Physics
and Astronomy Research Council (PPARC).}
	\hspace{0.2cm} (for the H1 and ZEUS Collaborations)\\
        School of Physics and Astronomy, University of Birmingham,
B15 2TT, UK.\\
        E-mail: \email{prn@hep.ph.bham.ac.uk}}

\abstract{Recent measurements of the diffractive cross section in
deep-inelastic scattering (DIS) at HERA are presented. The data are used to
investigate the factorisation properties of diffractive DIS and to
examine its quantum chromodynamic (QCD) structure. Models based on the 
colour dipole approach to DIS are also tested.}

\begin{document}

\section{Diffractive Deep Inelastic Scattering}

\EPSFIGURE[r]{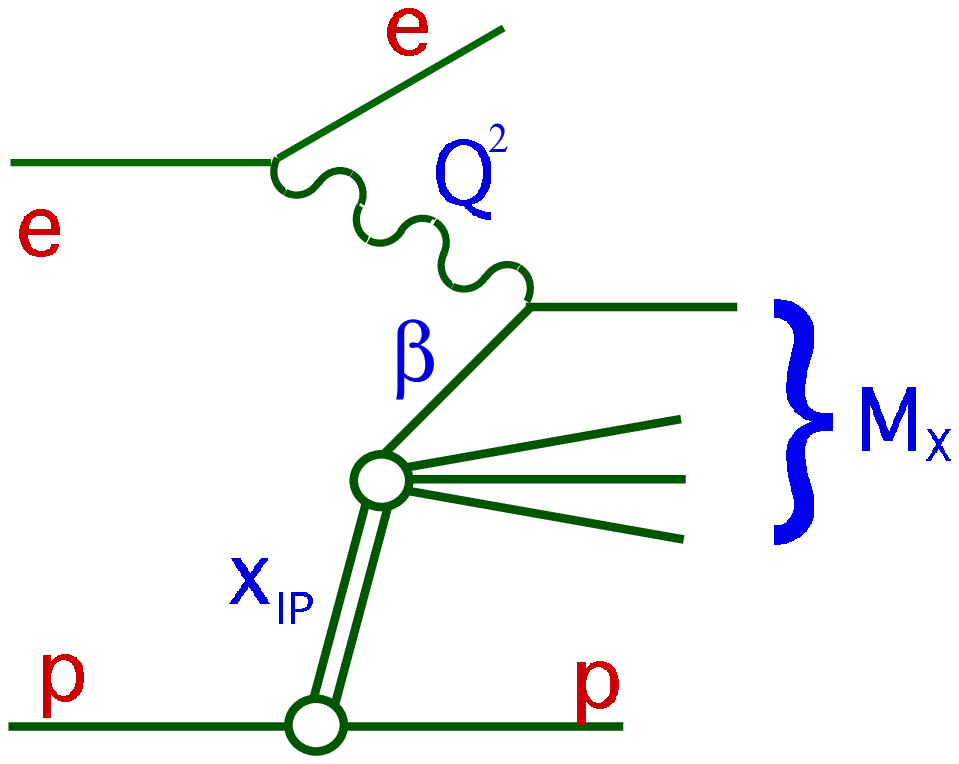,width=0.3\textwidth}{Illustration of the 
kinematic variables used to describe diffractive DIS.\label{kinematics}}

At low $x$ 
in DIS at HERA, approximately 10\% of the events are of the
type $e p \rightarrow eXp$, where the
final state proton carries in excess of 95\% of the proton beam 
energy \cite{h1:f2d94,zeus:f2d94}.
The kinematics of these processes are illustrated in figure~\ref{kinematics}.
A photon of virtuality $Q^2$, coupled to the electron, 
undergoes a strong interaction with the proton to form a final state hadronic 
system $X$ (mass $\mx$) separated by a large rapidity gap from the
leading proton.
No net quantum numbers are exchanged. 
A fraction $\xpom$ of the proton longitudinal momentum is transferred to
the system $X$. The virtual photon couples to a quark carrying a fraction 
$\beta$ 
of the exchanged momentum. The squared four-momentum transfer at the 
proton vertex is denoted $t$.

Events with this `diffractive' topology are interpreted in Regge models
in terms of pomeron trajectory
exchange between the proton and the virtual photon.
The large photon virtualities encourage a perturbative QCD treatment
of the process.
However, the parton level interpretation is not obvious. In order to
generate an exchange with net vacuum quantum numbers, a minimum of
two partons must be exchanged in the $t$ channel. 

The differential cross section for diffractive DIS
is often presented in terms of a diffractive structure function
$F_2^{D(4)} (\beta, Q^2, \xpom, t)$, defined analogously to the 
inclusive proton structure function $F_2$.
Experimentally, two complementary
methods have been used to measure $F_2^D$. Recent measurements
in which the leading proton is measured in proton
spectrometers 
are described 
in \cite{florian}. In this contribution, data are presented for which
it is not required that
the leading proton is detected and the kinematics are reconstructed 
from the hadronic system $X$.
This latter method yields the better statistical precision,
but does not allow a measurement of $t$. The results are therefore
presented in the form of a structure function $F_2^{D(3)} (\beta, Q^2, \xpom)$,
corresponding to an integral of $F_2^{D(4)}$ over $t$. 

The H1 collaboration recently released new preliminary $F_2^{D(3)}$ 
data \cite{h1:eps} (see 
figures~\ref{q2dep} and~\ref{betadep})
based on a factor of 5 more luminosity than previous 
measurements.\footnote{The new H1
data are integrated over $|t| < 1 \ {\rm GeV^2}$ and include a small
contribution ($5-10\%$)
from processes in which the proton dissociates to a
system of mass less than $1.6 \ {\rm GeV}$.} 
In the following sections,
these data are used together with previous data from ZEUS and H1 to
test the factorisation properties of diffractive DIS and its
relationship to inclusive DIS.

\section{Factorisation Properties and Diffractive Parton Densities}

\EPSFIGURE[r]{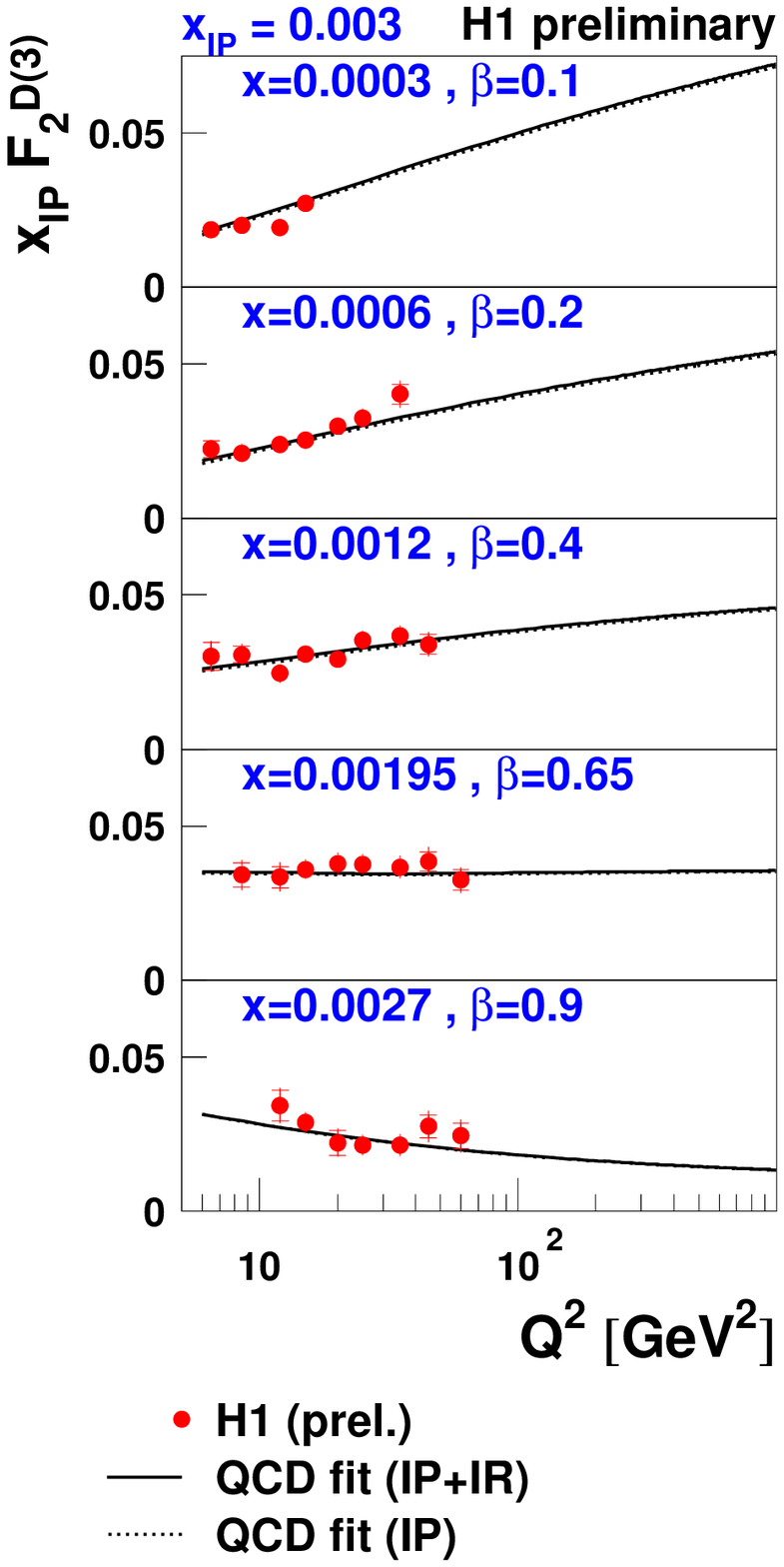,width=0.35\textwidth}{Dependence of 
$\xpom F_2^D$ on $Q^2$ for different $\beta$ values, with fixed
$\xpom = 0.003$. The data are compared with the DGLAP QCD fit described
in the text (from \cite{h1:eps}).\label{q2dep}}

In \cite{h1:eps}, the $\beta$ and $Q^2$ dependence of $F_2^{D(3)}$ 
is studied with high precision by measuring the structure function
at four fixed values of $\xpom = 0.001$, 0.003, 0.01 and 0.03. As 
examples, the results for $\xpom = 0.003$ are shown in
figures~\ref{q2dep} and~\ref{betadep}. 
In figure~\ref{q2dep}, scaling violations with positive 
$\partial F_2^D / \partial \ln Q^2$ persist up
to large values of $\beta > 0.4$, confirming earlier 
results \cite{h1:f2d94,zeus:f2d94}. Since 
$x = \beta \cdot \xpom$, the scaling violations in figure~\ref{q2dep} can be
compared with the scaling violations of the inclusive $F_2$ at the
same value of $x$. When compared at fixed $x$, the $Q^2$ dependences
of $F_2$ and $F_2^D$ are similar for $\beta \lapprox 0.65$ in the 
diffractive case, suggesting that
similar dynamics are at work in the two processes. 
At the highest $\beta$, the
logarithmic 
$Q^2$ derivative of $F_2^D$ becomes negative and there is a clear 
difference between the 
inclusive and diffractive $Q^2$ dependences at the same $x$ \cite{h1:eps}. 
In this high $\beta$ region,
higher twist contributions such as elastic vector meson production
are thought to play a major role in the diffractive cross 
section \cite{kgbw}.
The $\beta$ dependence of $F_2^D$ 
(figure~\ref{betadep}) is relatively flat. 

In \cite{collins}, hard scattering factorisation
was proven for a general class of semi-inclusive processes in DIS.
A particular case is leading proton production
with specified values of $\xpom$ and $t$, corresponding to the final
states measured in diffractive DIS at HERA. 
The $x$ 
and $Q^2$ dependence of 
the leading twist component of diffractive DIS can thus be treated in 
an analogous way to inclusive DIS. 
`Diffractive parton densities' of
the proton can be defined, which evolve according to the DGLAP equations 
and can be used to calculate observable cross sections when combined with 
suitable coefficient functions. 

\EPSFIGURE[r]{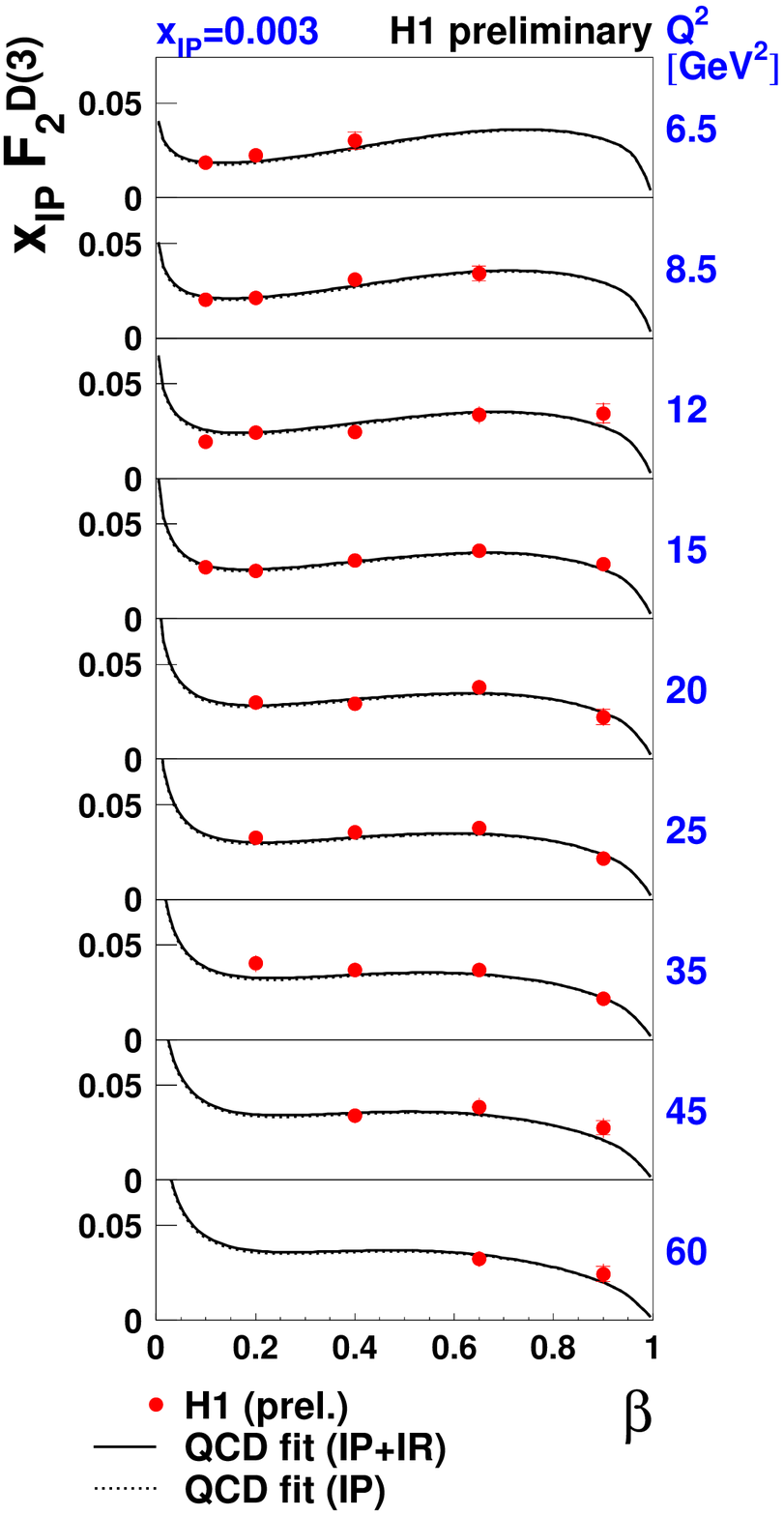,width=0.35\textwidth}{Dependence of 
$\xpom F_2^D$ on $\beta$ for different $Q^2$ values, with fixed
$\xpom = 0.003$. The data are compared with the DGLAP QCD fit described
in the text (from \cite{h1:eps}).\label{betadep}}

In figures~\ref{q2dep} and~\ref{betadep}, the data are compared with the
results of a fit in which the 
($\beta$, $Q^2$) dependence is obtained by parameterising
the diffractive light quark and gluon densities at $Q_0^2 = 2 \ {\rm GeV^2}$
and evolving to higher $Q^2$ using the leading order
DGLAP equations. 
The
$\xpom$ dependence is assumed to factorise from the ($\beta, Q^2$)
dependence and is  
described by a Regge phenomenological flux factor such that 
\begin{eqnarray}
\hspace*{-0.45cm} \xpom F_2^{D(3)} = A(\beta, Q^2)
\xpom^{2 - 2 \av{\alphapom(t)}} \sim
x^{2 - 2 \av{\alphapom(t)}} ,
\label{f2dregge}
\end{eqnarray}
where $\alphapom(t)$ is the effective pomeron trajectory.
The fit describes the data well and results in 
diffractive parton densities dominated by the
gluon density, which extends to large fractional momenta. 
Similar diffractive parton densities extracted from previous data 
have been highly successful in describing hadronic 
final state measurements in diffractive DIS \cite{hera:hfs}.

The hard scattering 
factorisation proof \cite{collins} makes no prediction for the
($\xpom$, $t$) dependence. From the QCD
perspective, the diffractive parton densities
could vary in both shape and normalisation with these variables.
However, the success of Regge phenomenology in describing soft
hadronic cross sections with a universal pomeron 
trajectory suggests that there
may be an extended `Regge'
factorisation property whereby the $\xpom$ 
dependence is driven by Regge asymptotics and is completely decoupled
from the ($\beta$, $Q^2$) dependence. The dependence on ($\beta$, $Q^2$) 
then represents a structure
function for the exchanged pomeron \cite{is}.

\EPSFIGURE[hr]{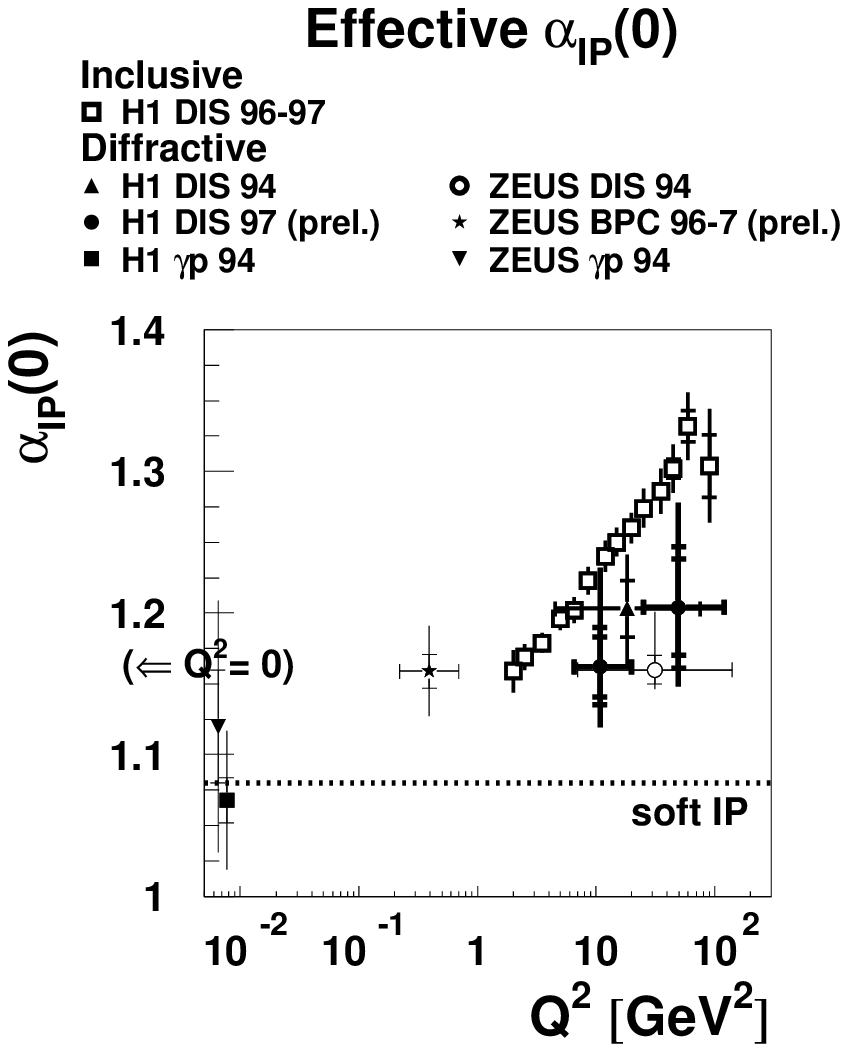,width=0.4\textwidth}{Compilation of values
extracted for the effective pomeron
intercept in inclusive and diffractive 
$ep$ scattering, shown as a function of 
$Q^2$.\label{alphapom}}

In \cite{h1:eps}, the Regge factorisation hypothesis 
is tested by measuring the data at a larger number of $\xpom$
values\footnote{$F_2^D$ is measured at a total of 312 points in the 
($\beta, Q^2, \xpom$) phase space.}
and performing a fit to equation (\ref{f2dregge}) 
with free parameters for the effective pomeron intercept $\alphapom(0)$
and $A(\beta,Q^2)$ at each ($\beta$, $Q^2$) point. At 
large $\xpom$ (equivalently small 
$\gamma^* p$ centre of mass energy $W$), 
contributions from sub-leading
exchanges are required\footnote{The fit yields a $\chi^2$ of 0.95 (1.25) per
degree of freedom with (without) a sub-leading term included.} 
in order to obtain a good fit to the data,
although the normalisation and effective intercept of this contribution
is not well constrained. The fit yields 
$\alphapom(0) = 1.173 \pm 0.018 \ ({\rm stat.}) \
\pm 0.017 \ ({\rm syst.}) \ ^{+ 0.063}_{-0.035} \ ({\rm model})$,
the dominant upward model dependence uncertainty arising from 
the unknown contribution of the cross section for longitudinally polarised 
photons. The Regge factorisation hypothesis works well within the 
kinematic range measured in \cite{h1:eps}, 
with no significant variation of the
effective $\alphapom(0)$ with $\beta$ or $Q^2$.
There is thus
no experimental evidence at the present level of precision 
for a variation of the diffractive
parton densities with $\xpom$. 
The measured $\alphapom(0)$ is compared with previous DIS 
and photoproduction measurements and a recent ZEUS measurement in 
the low $Q^2$
transition region \cite{zeus:loq} in figure~\ref{alphapom}.
The result for diffractive DIS is significantly larger than
that describing soft hadronic and photoproduction cross sections \cite{dl}.

Simple Regge predictions for the total $\gamma^* p$ cross section
lead at fixed $Q^2$ to
\begin{eqnarray}
F_2(x, Q^2) \propto x^{1 - \alphapom(0)} \ .
\label{f2regge}
\end{eqnarray}
Figure~\ref{alphapom} also shows the effective pomeron intercept extracted
from inclusive DIS
at low $x$ using equation (\ref{f2regge}) \cite{lambda}. The
effective intercepts 
describing the inclusive and diffractive energy dependences become 
different at large $Q^2$. From equations (\ref{f2dregge}) and
(\ref{f2regge}), 
Regge pole models predict
a factor of approximately 2 difference
in the power of the growth 
of the diffractive and inclusive cross sections
with decreasing $x$ (increasing $W$). Experimentally,
the ratio of diffractive to inclusive cross 
sections in DIS is found to be relatively flat as a function of $x$
when $\mx$, $\beta$ and $Q^2$ are fixed \cite{zeus:f2d94,h1:eps}. 
The situation for 
the low $Q^2$ transition region
is rather
different \cite{zeus:loq}, 
since the Regge predictions for the ratio of diffractive
to inclusive cross sections work well.

\section{Comparisons with Dipole Models}

The hard scattering factorisation proof for diffractive DIS 
does not specify the relationship between the diffractive and the
inclusive parton densities. Specific models 
(e.g. \cite{bgh,kgbw}) have been 
developed for this relationship. A popular approach
is to consider the interaction in the proton rest frame, in
terms of the elastic and total cross sections for the scattering 
on the target of
$q \bar{q}$ and $q \bar{q} g$ fluctuations of the 
virtual photon, treated
as colour dipoles. 
Using ideas such as the optical theorem,
the same `dipole cross section' 
can be used to describe total, elastic
and dissociative cross sections, thus unifying the description of 
$F_2$ and $F_2^D$.
As yet, there is no consensus on the
proper way to treat the dipole cross section.

\EPSFIGURE[r]{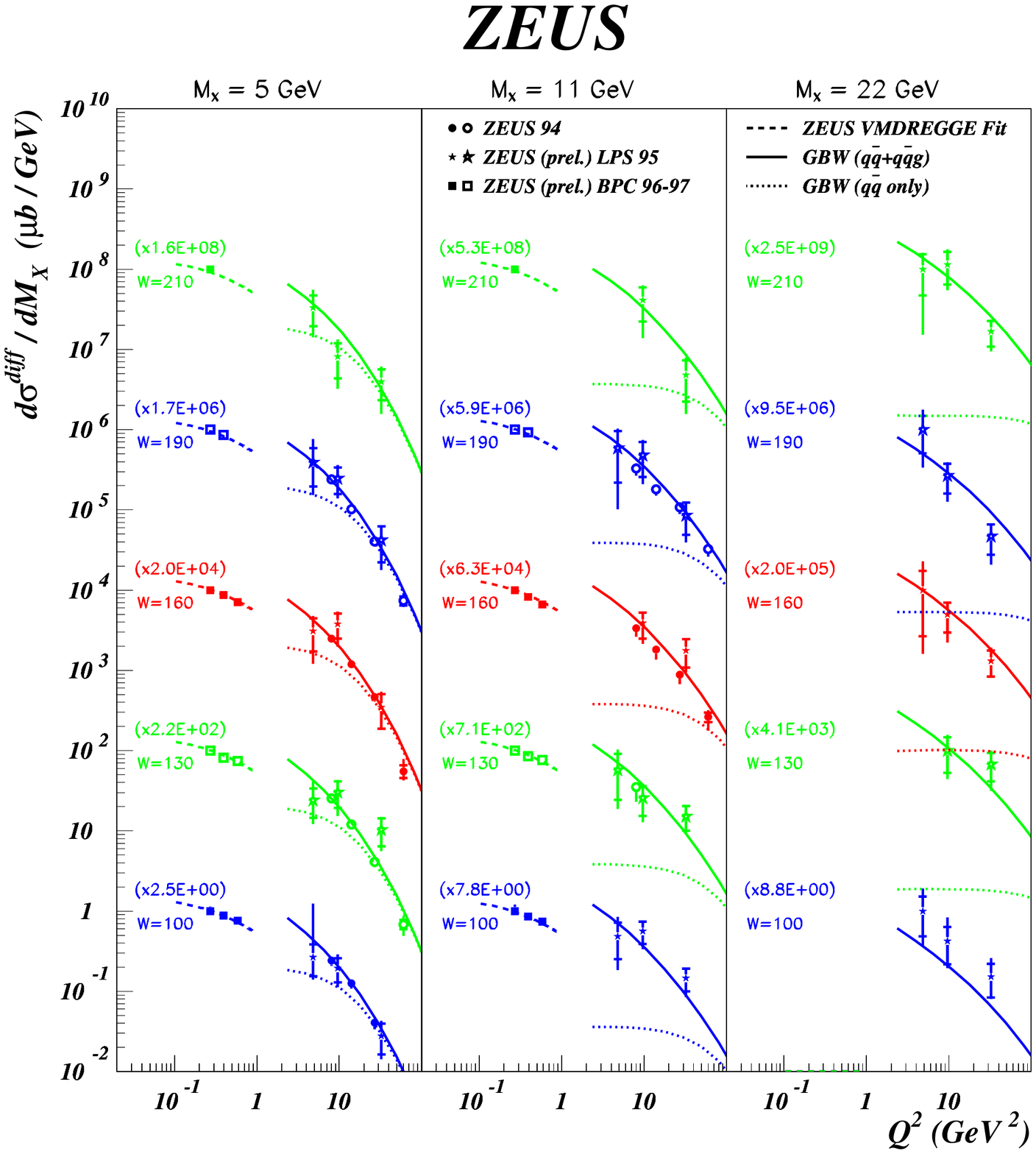,width=0.59\textwidth}{Compilation of ZEUS $F_2^D$
data, compared at high $Q^2$ with the ``saturation'' model \cite{kgbw} 
and at low $Q^2$ with a Regge motivated parameterisation.\label{satn}}



In the ``saturation'' model \cite{kgbw}, the $q \bar{q}$ dipole 
cross section is obtained from 
a 3 parameter fit to $F_2$ data and 
is then used to 
predict $F_2^D$,
under the assumption that the diffractive cross section is driven by
2-gluon exchange.\footnote{The prediction is for $F_2^{D(4)}$ at $t=0$.
To describe the $F_2^{D(3)}$ data, an additional free parameter is needed,  
corresponding to the exponential $t$ dependence, parameterised
as $e^{Bt}$.}
A contribution from $q \bar{q} g$ fluctuations is added in the diffractive
case, assumed to interact with 
the same dipole cross section as the $q \bar{q}$ fluctuation.
Figure~\ref{satn} shows a comparison 
of the ``saturation'' model 
with various diffractive data from ZEUS \cite{zeus:f2d94,zeus:loq}. 
The description is good for 
$Q^2 \geq 4 \ {\rm GeV^2}$. The $q \bar{q} g$ contribution
is clearly needed at large $\mx$. As yet, the model is not able
to describe the low $Q^2$ region.

Considering the small number of parameters, the model in \cite{kgbw} gives 
a good description of the new H1 data in \cite{h1:eps}, though
there are clear discrepancies in the small $\beta$, small $Q^2$ region. 
Including QCD evolution of the gluon distribution \cite{kgbnew} 
does not improve the
description of the data.

In the
``semi-classical'' model \cite{bgh}, the dipole cross section is
modelled as the scattering from a superposition of colour fields
of the proton according to a simple non-perturbative model.
All resulting final state configurations contribute to the inclusive proton
structure function. Those in which the scattered partons 
emerge in a net colour-singlet state
contribute to the diffractive structure
function. 
The model contains only four free parameters,
which are obtained from a combined fit to previous $F_2$ and $F_2^D$
data.
The model reproduces the general features of the $F_2^D$ data in \cite{h1:eps},
but also lies above the data where $\beta$ and
$Q^2$ are both small.  



\end{document}